\documentclass[5p,twocolumn]{elsarticle}

\usepackage[numbers]{natbib}
\usepackage{lineno}
\usepackage{epstopdf}
\usepackage[official]{eurosym} 
\usepackage{amsmath}
\usepackage[utf8]{inputenc}
\usepackage[T1]{fontenc}
\usepackage{booktabs}  
\usepackage{multirow}
\usepackage{stfloats}    
\usepackage{rotating}				
\usepackage{textcomp}			
\usepackage{fancybox}
\usepackage{bigstrut}
\usepackage{tabularx}
\usepackage{optidef}		

\usepackage{cuted}

\PassOptionsToPackage{hyphens}{url} 
\usepackage[breaklinks=true]{hyperref}
\newcolumntype{L}[1]{>{\raggedright\arraybackslash}p{#1}} 
\newcolumntype{C}[1]{>{\centering\arraybackslash}p{#1}} 
\newcolumntype{R}[1]{>{\raggedleft\arraybackslash}p{#1}} 

\journal{Energy}

\biboptions{sort&compress}

\begin{document}

\begin{frontmatter}

	\title{Future competitive bioenergy technologies in the German heat sector: Findings from an economic optimization approach}

	\author[UFZ]{Matthias Jordan\corref{cor1}}
	\ead{matthias.jordan@ufz.de}
	
	\author[DBFZ]{Volker Lenz}
	\author[UFZ]{Markus Millinger}
	\author[DBFZ]{Katja Oehmichen}
	\author[UFZ,DBFZ]{Daniela Thrän}
	
	\address[UFZ]{Helmholtz Centre for Environmental Research - UFZ, Permoserstraße 15, 04318 Leipzig, Germany}
	\address[DBFZ]{DBFZ Deutsches Biomasseforschungszentrum gGmbH, Torgauer Strasse 116, 04347 Leipzig, Germany}
	
	\cortext[cor1]{Corresponding author}
	




	\begin{abstract}

Meeting the defined greenhouse gas (GHG) reduction targets in Germany is only possible by switching to renewable technologies in the energy sector. A major share of that reduction needs to be covered by the heat sector, which accounts for $\sim35\%$ of the energy based emissions in Germany. Biomass is the renewable key player in the heterogeneous heat sector today. Its properties such as weather independency, simple storage and flexible utilization open up a wide field of applications for biomass. However, in a future heat sector fulfilling GHG reduction targets and energy sectors being increasingly connected: which bioenergy technology concepts are competitive options against other renewable heating systems? In this paper, the cost optimal allocation of the limited German biomass potential is investigated under long-term scenarios using a mathematical optimization approach. The model results show that bioenergy can be a competitive option in the future. Especially the use of biomass from residues can be highly competitive in hybrid combined heat and power (CHP) pellet combustion plants in the private household sector. However, towards 2050, wood based biomass use in high temperature industry applications is found to be the most cost efficient way to reduce heat based emissions by 95\% in 2050.
	\end{abstract}

	\begin{keyword}
	\texttt heat sector \sep bioenergy  \sep renewable energy \sep optimization \sep hybrid heat technologies
	\end{keyword}

\end{frontmatter}


\section{Introduction}

Global climate change, depleting energy resources and energy security are issues affecting all countries. In Germany ambitious emission reduction and efficiency improvement targets are defined by the government \cite{Bundesregierung.2010}. GHG emissions are to be reduced by $80-95\%$ until 2050 compared to 1990 by improving efficiency and switching to renewable technologies in the energy sector. A major share of that reduction needs to be covered by the heat sector, which accounts for $\sim35\%$ of the energy based emissions \cite{Umweltbundesamt.2017c} and 54\% of the final energy demand \cite{BundesministeriumfurWirtschaftundEnergie.2018} in Germany today.

The German heat sector is characterized by its heterogeneity due to different demand profiles, applications and infrastructures. Heat consumption takes place in millions of residential buildings (which accounts for 43\% of the final heat demand), trade and commerce buildings (17\%), as well as in many different fields of the industry (40\%) \cite{BundesministeriumfurWirtschaftundEnergie.2018}, mainly the steel and chemical industries in high temperature applications. Within these sectors, different temporal demands occur, ranging from seasonal to daily fluctuating needs. In addition to this complex demand structure, 8\% of heat is not produced at the location of demand, but distributed via district heating grids \cite{BundesministeriumfurWirtschaftundEnergie.2018}. To reduce greenhouse gas emissions in the heat sector both the demand and supply sides need to be addressed.

Heat demand in buildings needs to be decreased by increasing the refurbishment rate. Additionally, the heat transition needs different renewable technological solutions that fit this complex market structure, combining renewable power and biomass energy sources.

In 2017, biomass was the largest renewable energy contributor in Germany (54\%), particularly in the heat sector where 87\% of the renewable energy was covered by biomass. Solid biomass was contributing the highest share of renewable heat with 68\% \cite{Umweltbundesamt.2018}. However, alternative renewable heat options take up more market shares, the resource biomass is limited and a great share of the German yearly usable potential is already exploited \cite{Brosowski.2015}. On the other hand, bioenergy has clear advantages compared to other renewable fluctuating energy sources in the heat sector: weather independency, the possibility of simple storage and flexible utilization. These properties open up a wide field of application for biomass within the different sub-sectors of the heat sector. But in which sub-sectors is biomass competitive against other renewable applications, while fulfilling the GHG reduction targets?

Several studies are available on the development of the German energy transition in general \cite{Nitsch.2012,Repenning.2015,Schlesinger.2014,Pfluger.2017}, focusing on the power sector and examining energy from biomass only roughly. \citet{Thran.2015b} investigated the allocation of biomass in different German energy sectors. The results show that wood based biomass in the transport and power sector is only competitive under special circumstances, expecting to have more competitive applications in the heat sector, which was not modelled in the mentioned study. To the authors' knowledge, there is no study modelling the complex structure of the complete heat sector in detail, while including hybrid heating technologies and representative bioenergy technology concepts, also in combination with other renewable technologies. Additionally, reviews focussing on model-based analysis in the heat sector, do not identify any studies combining the above mentioned research intentions \cite{Bloess.2018,Merkel.2017}.

In this paper, the cost optimal allocation of biomass between different heat sub-sectors is investigated in the frame of long-term energy scenarios.  The following research question is assessed:

-	Which bioenergy technology concepts are competitive options in a future, climate target fulfilling heat sector and how does their potential role differ in different heat sub-sectors?

\section{Materials and method}

In this study, the heat sector was divided into several sub-sectors, with different properties in terms of demand profiles and infrastructures. Representative bioenergy-, fossil- and other renewable (hybrid-)heat-technology concepts were defined for each sub-sector and the technological competition was optimized in the system within the framework of the German climate protection plan \cite{BundesministeriumfurUmweltNaturschutzBauundReaktorsicherheit., Bundesregierung.2010} in two scenarios. A consistent scenario framework was set up and detailed biomass feedstock data were defined, leading to a set of five biomass types, which can be processed into 20 biomass products. With additionally three fossil products, they can be applied to 47 different technology concepts. Within the model these technology concepts were in competition on 19 different sub-sectors to identify the optimal allocation of biomass in the heat sector.

\subsection{Modelling}
\label{sec:Modelling}

A mathematical optimization approach was chosen to model the heat sector. The approach of the model follows BENOPT (BioENergyOPTimisation model), which has been applied on the transport and power sector \cite{Millinger.2018,Millinger.2019,Millinger.2019b}. As a programming environment GAMS \cite{GAMSDevelopmentCorp..2019} is used in combination with MATLAB \cite{TheMathWorks.2019}. GAMS is an algebraic modelling language for mathematical optimization. In Matlab the input data is imported from Microsoft Excel \cite{Microsoft.2019}, edited and automatically sent to GAMS, where the minimum costs are calculated. The results from the optimizer are exported back to Matlab, where they are evaluated and graphically prepared.

The model in this paper is fully deterministic and uses perfect foresight. The technology choice is optimized within the competition. It is a linear model, using the Cplex solver. The spatial boundary is Germany as a whole. The objective function is minimizing the total system costs over all technologies \textit{i}, all sub-sectors \textit{s} and the complete timespan \textit{t}=2015...2050 \eqref{eq:obj}. The total system costs are the sum of the technology specific marginal costs \textit{mc}, multiplied with the amount of heat produced \textit{$\pi$}, and the investment costs \textit{ic}, discounted with the annuity method (discount rate $q$) \cite{Heuck.2010}, multiplied with the number of heating systems installed \textit{$n^{cap}$}. In the model each (hybrid-)heat-technology concept is separated into different modules \textit{j}, assigned with different lifetimes \textit{$\hat{t}$} and individual investment costs.

\begin{flalign}	
	\intertext{\textbf{Objective function}}
	\begin{split}
		min\sum_{t,i,s,b}mc_{t,i,s,b}\cdot \pi_{t,i,s,b}& \\ +\sum_{t,i,j,s}ic_{t,i,j,s}\cdot n^{cap}_{t,i,j,s} \cdot\frac{q(1+q)^{\hat{t}_{j}}}{(1+q)^{\hat{t}_{j}}-1}&
	\end{split}\label{eq:obj} \\[2ex]
	\intertext{\textbf{subject to}}
	\delta_{t,s}=\sum_{i,b} \pi_{t,i,s,b},\forall (t,i,s,b) \in (T,I,S,B)& \\[2ex]
	\begin{split}\phi^{Res}_{t}+\Lambda^{Land}_{t}\cdot Y_{t,b}\geq\sum_{i,s,b}\dot{m}_{t,i,s,b},& \\ \forall (t,i,s,b) \in (T,I,S,B_{bio})&\end{split} \\[2ex]
	\begin{split}\varepsilon^{max}_{t}\geq\sum_{i,s,b}\alpha_{i,s}\cdot(\varepsilon^{rel}_{t,i,s}\cdot \pi_{t,i,s}+\varepsilon^{feed}_{t,i,s,b}\cdot \dot{m}_{t,i,s,b})  	\label{eq:GHG_em},& \\ \forall (t,i,s,b) \in (T,I,S,B)& \end{split} \\[2ex]
	 \pi_{t,i,s,b}=\dot{m}_{t,i,s,b}\cdot \eta_{t,i,s},\forall (t,i,s,b) \in (T,I,S,B)&\label{eq:PRD_BC} \\[2ex]
	n^{cap}_{t=2015,i,j,s}=n^{initial}_{i,j,s},\forall (t,i,j,s) \in (T,I,J,S)&\label{eq:N1} \\[2ex]
	\begin{split}n^{cap}_{t+1,i,j,s}=n^{cap}_{t,i,j,s}+n^{ext}_{t+1,i,j,s}-n^{dec}_{t+1,i,j,s},& \\ \forall (t,i,j,s) \in (T,I,J,S)&\end{split} \\[2ex]
	\begin{split}n^{dec}_{t,i,j,s}=n^{initialdec}_{t,i,j,s}+n^{extdec}_{t,i,j,s},& \\ \forall (t,i,j,s) \in (T,I,J,S)&\end{split} \\[2ex]
	n^{extdec}_{t+\hat{t}_{j},i,j,s}=n^{ext}_{t,i,j,s},\forall (t,i,j,s) \in (T,I,J,S)&\label{eq:N4}
\end{flalign}

 Marginal costs include feedstock costs (fossil or biomass), costs for power demand, maintenance and a CO$_2$-certificate price. The sum of these costs has a dynamic development, which depends on the time point, used technology, sub-sector and if applicable the consumed feedstock product \textit{b}. Generated power in a combined heat and power (CHP) system is included as a credit within the variable costs. For details on how the credit is calculated see section \ref{sec:SectorCoupling}.

The main model restrictions are as follows: First, the heat demand $\delta$ in each sub-sector needs to be fulfilled. Therefore the sum of the produced heat within one sub-sector equals the heat demand within a sub-sector in each year. Second, the yearly consumed biomass $\dot{m}$ within the system must not be higher as the sum of the limited biomass potential from residues $\phi^{res}$ and the limited land use potential $\Lambda^{Land}$ multiplied with the corresponding yield $Y$ of the energy crop. More details on the biomass potential and possible biomass pathways are explained in section \ref{sec:FeedstockData} and \ref{sec:Scenarios}. Third, the yearly maximal allowed amount of GHG emissions $\varepsilon^{max}$, representing the federal climate targets in Germany, must be greater or equal to the sum of the technology-based $\varepsilon^{rel}$ and feedstock-based $\varepsilon^{feed}$ emissions \eqref{eq:GHG_em}. The relationship between the produced heat and the utilised feedstock product is given in equation \eqref{eq:PRD_BC} and determined by the conversion efficiency $\eta$ of each technology. Equation \eqref{eq:N1} to \eqref{eq:N4} explain the relationship between the number of heating systems installed ({$n^{cap}$) at time point \textit{t}, the number of heating systems newly invested in ($n^{ext}$) and the number of heating systems decommissioned ($n^{dec}$). The status quo of all installed heating systems in 2015 serves as a starting point ($n^{initial}$). This portfolio is linearly decommissioned over the corresponding lifetime of each technology ($n^{initialdec}$). Heating systems newly installed in the model ($n^{ext}$) are decommissioned after they have reached their lifetime, defined by the variable $n^{extdec}$. Premature decommissioning of heating systems is only allowed for fossil technologies and limited to 1\%/a. As a restriction for energy crops, every type may maximally double its land use per year.

\subsection{Heat sub-sectors}
\label{sec:Markets}

Heat utilisation differs from power utilisation, which is supplied through one uniform grid with a unique frequency and different voltage levels which can be transformed up and down. For heat supply, beside local heating grids, differing in temperature, pressure and extension, numerous single object solutions exist, with temperatures ranging from 1.000 \textdegree C for industrial processes down to low temperature heating with about 40 \textdegree C \cite{Umweltbundesamt.}. Additionally, the amount of heat required differs, with a corresponding capacity variation for heat generators. Furthermore, heating systems based on solid fuels (biomass, coal or waste) vary in terms of operation efficiency and emissions depending on the load \cite{Kaltschmitt.2016}. Differing patterns for peak demand, yearly demand variations, temperature requirements and the relation between base load (e.g. hot water supply) and the varying proportion of the heat demand (e.g. space heating) require specially adapted technology concepts. Thus, heat demand can be divided into a whole series of sub-sectors in which different heating concepts have to be applied.

In reality, each heating object is individually examined and a decision on the best case is taken by the owner or an ordered decision maker according to an individual set of decision parameters and the knowledge of the involved actors. For an artificial model, a fixed set of decision parameters is required as well as a simplification of the decision cases (see section \ref{sec:Modelling}). Therefore, similar demand cases were aggregated to one sub-sector with mean values and a certain set up of suitable technology options. Special cases with low heat demands were included in the most suitable sub-sector.

The main difference in the heat supply depends on the required temperature level, which is basically distinguished between industrial applications (60 °C to more than 1.000 °C) and building heat demand (usually less than 95 °C). Considering comparable renewable heating concepts, industrial heat supply was separated into four sub-sectors by different temperature levels \cite{Kemmler.2017}:\newline< 200 °C, 200 - 500 °C, 500 - 1.500 °C and one sub-sector for special coal demand (fossil or bio-coal) in industrial applications for steel production.

In addition to industrial applications, more than $50\%$ of the total heat demand in Germany is used for space heating and hot water supply at a temperature level below 95 °C \cite{Umweltbundesamt.}. When supplying individual objects of different sizes with fossil systems, no major technological difference is required. A heat supply by bioenergy, however, requires the use of different technological solutions depending on the size of the boiler. From smaller applications in single family houses using stoves or wood log boilers, through pellet boilers in multi-family houses up to wood chip boilers in e.g. schools or hospitals, a variety of technological solutions and combinations are possible \cite{Kaltschmitt.2016}. Additionally, CHP-technologies based on solid biomass fuels are favourable options for cases with a high base load demand, such as in indoor swimming pools. Considering these aspects, the private household and trade and commerce sector was structured into 14 sub-sectors according to the peak demand, the relation of hot water demand to total heat demand and the required temperature levels \cite{Lenz.2019}. The future development of the heat demand in each sub-sector is based on the external results of the model 'B-STar' \cite{Koch.2018}. As a stocks exchange model, it represents the building stock in Germany and models the future refurbishment in different scenarios.

Centralized heating supply was summarized in one sub-sector, determined by the resolution of the data basis. 

In total, 19 sub-sectors were defined and described (see \citet{Lenz.2019} and Table \ref{tab:Tech2Market} in appendix B). The average thermal peak load demand and the annual final heat demand until 2050 serve as input data for the optimization model and the design of the different technology concepts in each sub-sector.
	
\subsection{Technology concepts}
\label{sec:Technologies}

In order to determine the future use of biomass in the heat sector, the market competition has to be depicted in the optimization model. Consequently, different fossil and renewable technological systems were selected for the competition in each sub-sector. Beside single technology solutions, also hybrid systems were included. Hybrid systems are combining different types of fuels, leading to a variety of possible technical solutions. For the final selection of the defined heating concepts, the following aspects were taken into account: 

\begin{itemize}
	\item The status quo of the national biomass feedstock mix and all installed heating systems in 2015 were considered.
	\item As the research is focused on biomass, at least one bioenergy heat concept as well as one bioenergy CHP concept, based on solid fuels, is integrated in each sub-sector.
	\item Solar thermal was integrated as an established technology on the market.
	\item One heat pump concept per low temperature sub-sector was defined, as this technology offers the potential to fulfil the complete heat demand for applications lower than 200 °C in a renewable way.
	\item In order to ensure a net renewable power supply for heat pumps, a heat pump concept is always designed in combination with a PV system, which produces the major share of the electricity demand over the year.
\end{itemize}

As the most competitive fossil references a gas boiler or gas boiler in combination with a solar thermal system as well as a gas fuel cell plus solar thermal system were defined in the most cases. Oil-fired boilers were not included in the modelling as they are more costly and emit more CO$_2$ equivalents than gas-fired boilers. Every gas-fired concept can either obtain natural gas or biomethane, which is fed into the gas network. Different single bioenergy solutions were described according to the amount of heat and the thermal peak demand. Additionally, bioenergy hybrid or multibrid systems including a heat-pump, solar thermal or PV were selected according to the heat demand parameters of the sub-sector. Future technical improvements were considered through yearly increase rates of thermal efficiency, electrical efficiency and a decrease in investment costs \cite{Lenz.2019}. For gasification systems, a change from combustion engines to fuel cells is considered within the next two decades.

Table \ref{tab:Tech2Market} and \ref{tab:Tech2MarketIndustry} in appendix B show which concepts are considered in which sub-sectors. As there are some basic differences in the concepts between heating in buildings and industrial/ district heating provisions, these two sectors are shown in separate tables. However, the allocation of biomass over the sub-sectors is treated equally.

In total, 42 technical concepts where described. The complete technical and economic data for each technology concept per sub-sector can be found in a published data set \cite{Lenz.2019}. The calculated infrastructure emission factors of the single technology components as well as the feedstock specific emission factors are attached in table \ref{tab:EmissionFactor} and \ref{tab:FeedstockEmission} of appendix B.

\subsection{Feedstock data}
\label{sec:FeedstockData}

According to the above described technology concepts, four main feedstocks are considered in this model to generate heat or combined heat and power. Biomass from residues and energy crops is used for all bioenergy technologies. The basis for all other renewable heat technologies is the usage of electricity and for the most competitive fossil technologies gas and coal have been chosen as a reference. The heat production from plastic waste has been set as a constant to the amount of generation in 2015. Details on fossil and power based energy prices are shown in Fig. \ref{fig:PowerGasPrice}.

The technical potential for biomass residues are shown until 2050 based on \citet{Brosowski.2015}, shown in Fig. \ref{fig:BiomassPotential}. Additionally, crops for energetic and material use are cultivated on 2.4 Mio ha of land in Germany today \cite{Becker.2018b}. In this study, the maximum permitted land use is reduced linearly to 2.0 Mio ha in 2050, which is at the lower limit of identified values from currently available long-term energy scenario studies \cite{Nitsch.2012,Repenning.2015,Schlesinger.2014,Pfluger.2017}. On this land area, ten types of energy crops are cultivated for heat and CHP applications today \cite{Becker.2018}. In table \ref{tab:YieldLandUse} of appendix B the applied yields and the status quo of land use for these crops in the year 2015 are attached.

\begin{figure}[h]
	\centering
		\includegraphics[trim = 5px 10px 10px 5px, width=0.45\textwidth]{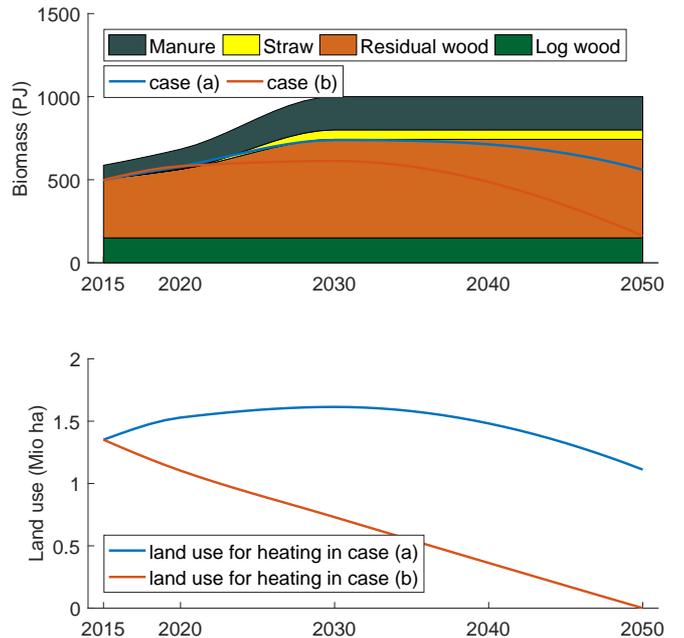}
		\caption{Technical biomass potential from residues in Germany \cite{Brosowski.2015} (top). Available pre-allocated biomass potential and available land area in case (a) and (b) shown by the coloured lines. The model is free to pick from any category of residues and is free to cultivate any of the defined energy crops, as long as the defined upper scenario limit is not violated.}
	\label{fig:BiomassPotential}
\end{figure}

\begin{figure}[ht]
	\centering
		\includegraphics[trim = 5px 10px 10px 5px, width=0.45\textwidth]{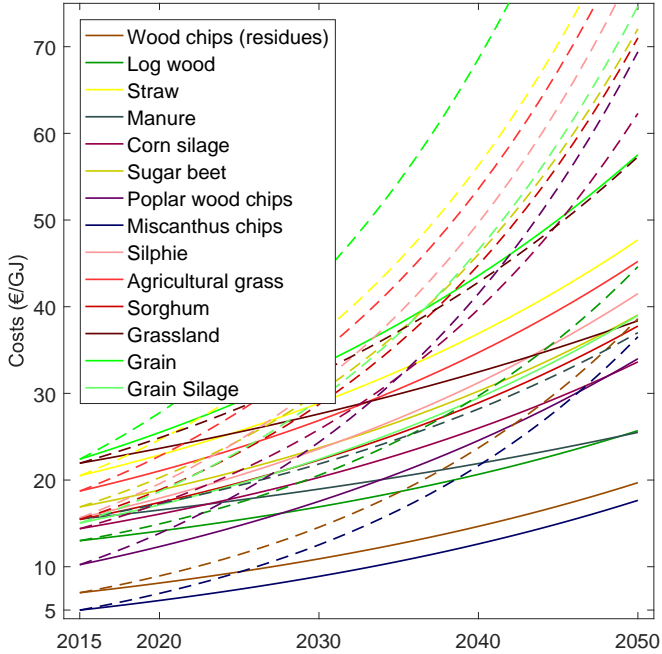}
		\caption{Cost developments of the biomass feedstocks for a yearly wheat price increase of 3\% (solid lines) and 5\% (dotted lines).}
	\label{fig:FeedCost}
\end{figure}

Different prices arise for the defined feedstocks. A common method to estimate future prices of energy crops is to add the per hectare profit of a benchmark crop to the per hectare production costs of the energy crops \cite{Witzel.2016}. In Germany, the most common crop is wheat \cite{StatistischesBundesamt.2016}, which holds for the benchmark crop in this study. Based on the price increase of wheat in the last decades \cite{WorldBank.2019}, two biomass price development scenarios are modelled in this study with a yearly increase of wheat by 3\% and 5\%. For a detailed description of the applied method in this paper the reader is referred to \citet{Millinger.2016}. Prices for biomass products from residues in 2015 are according current prices \cite{agrarheute.2018,C.A.R.M.E.N.e.V..2018,TFZ.2018}. For the future development, the yearly increase rate of wheat in the corresponding scenario is also applied to biomass residues. Fig. \ref{fig:FeedCost} shows the resulting price development of the considered biomass feedstocks. Applied surcharges for extra processing steps, such as pelletising etc. can be found in table \ref{tab:surcharges} of appendix B.

\begin{table*}[!b]
  \centering
  \caption{Model linkage of the heat sector to the power sector in terms of power consumed for heating and power use of CHP / PV technologies. The emissions from grid-based electricity are allocated to the heat sector in accordance to the power mix specific emission factor \cite{Repenning.2015}.}
	\begin{tabular}{rccc}
	\toprule
	\multicolumn{1}{c}{\textbf{Power}} & \textbf{Price} & \textbf{Credit} & \textbf{Heat sector emissions} \\
	\midrule
	\textbf{external demand} & Final consumer price & 0     & Emissions from grid power mix \\
	\textbf{internally used for heating} & 0     & 0     & Emissions from techn. system \\
	\textbf{internally used for non heating} & 0     & Final consumer price & 0 \\
	\textbf{fed into the grid} & 0     & Stock market price & 0 \\
	\bottomrule
	\end{tabular}%
  \label{tab:LinkagePower}%
\end{table*}%

Biomass from residues and energy crops can be converted into several secondary energy carriers. In this study, 20 biomass products and three fossil products have been defined. Table \ref{tab:BiomassProducts2Tech} in appendix B shows which products can be used in which technologies. All fermentable feedstocks are processed into biomethane, which is fed into the gas supply network. Since multiple options per technology are possible, a differentiation between feedstock specific and technology specific emissions has to be made. Table \ref{tab:EmissionFactor} and \ref{tab:FeedstockEmission} in appendix B give an overview of the technology and feedstock specific emission factors and the corresponding allocation factors applied. 

\subsection{Sector coupling}
\label{sec:SectorCoupling}

The heat sector is strongly linked to the power sector, especially when CHP and power to heat options are modelled. To generate conclusive results for the heat sector, a linkage to the power sector is inevitable. In order to achieve this linkage, a scenario framework was set up. Certain input parameters, such as the electricity price, the electricity-mix specific emission factor and the CO${_2}$ certificate price, which are highly influential for the market development of the heat sector, do also rely strongly on the development of the power sector. These parameters and predicted fossil feedstock price developments are adopted from the 'KS95' scenario of the study of \citet{Repenning.2015}. Governmental subsidies, such as e.g. the EEG are not considered in this study. The only market steering instrument is the CO${_2}$ price, which is applied on the complete heat sector. As a result, the linkage of the heat sector to the power sector in relation to power prices, feed-in tariffs, own electricity consumption and emission allocation is shown in Table \ref{tab:LinkagePower}.

\citet{Repenning.2015} projects the future development of power and gas prices for the energy only markets. The required end consumer prices for our investigations are calculated consumption-dependent according to the monitoring report of the federal network agency for the model starting year 2015 \cite{Bundesnetzagentur.2017}. The future price developments are projected combining both sources \cite{Bundesnetzagentur.2017,Repenning.2015}, see Fig \ref{fig:PowerGasPrice}.

\begin{figure}[ht]
	\centering
		\includegraphics[trim = 5px 10px 10px 5px, width=0.45\textwidth]{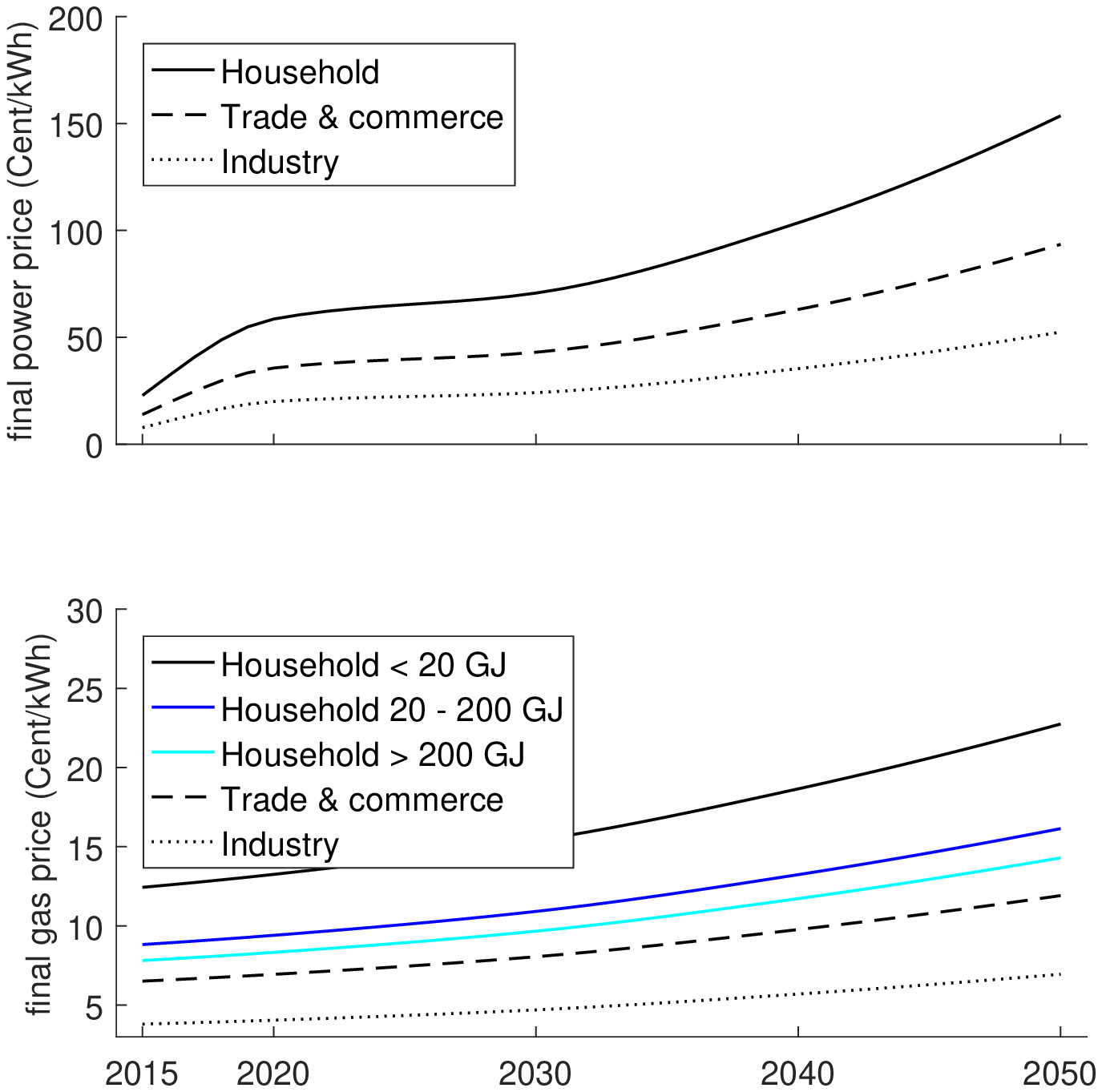}
		\caption{End consumer power (top) and gas (bottom) prices. Own calculations based on \citet{Repenning.2015} and \citet{Bundesnetzagentur.2017}.}
	\label{fig:PowerGasPrice}
\end{figure}

\subsection{Scenarios}
\label{sec:Scenarios}

In this study, a scenario of 95\% GHG emission reduction compared to 1990 is analysed. The focus of the investigation lies on the development of biomass in the heat sector, but still considering the interactions to other energy sectors by setting a scenario framework, derived from the 'KS95' scenario from the study of \citet{Repenning.2015}. From currently available long term energy scenarios in Germany \cite{Nitsch.2012,Repenning.2015,Schlesinger.2014,Pfluger.2017}, \citet{Repenning.2015} is the only one modelling a transformation path towards a 95\% reduction scenario and also reaching this target in 2050. However, within the study of \citet{Repenning.2015} biomass is depicted in a rough level of detail and only a minor share of the available biomass potential is distributed to the heat sector in the 'KS95' scenario. In this paper, a broader range of biomass potential is pre-allocated to the heat sector. \citet{Szarka.2017} reviews the role of bioenergy in long-term energy scenarios. The allocation of biomass to the heat sector in 2050 varies strongly between the reviewed studies, ranging from $\sim5 - 70 \%$ of the overall potential. 

\begin{figure*}[!b]
	\centering
		\includegraphics[trim = 40px 0px 40px 5px, width=0.95\textwidth]{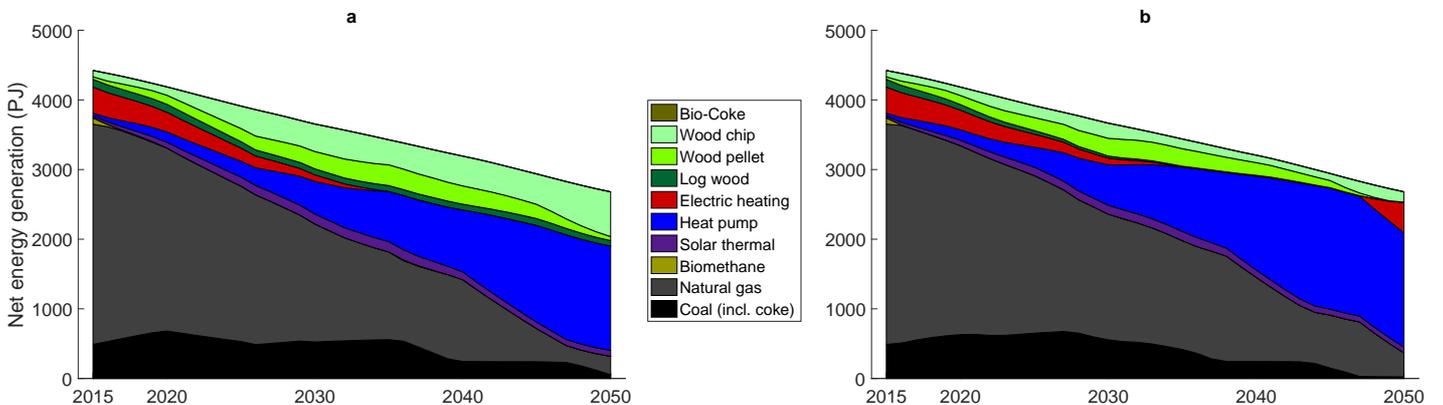}
		\caption{Model resulting development of the technology market shares for the complete heat sector in case (a) and (b) in a yearly resolution.}
	\label{fig:Result_Market}
\end{figure*}

Hence, two extreme scenarios are investigated in this paper, where one time a major share of the biomass potential (case a) and the other time a minor share of the biomass potential (case b) is pre-allocated for heating applications, for details see Fig. \ref{fig:BiomassPotential}. Consequently, the biomass potential for heat applications is fixed for each year and scenario, but the model is free to pick from any category of residues and is free to cultivate any of the defined energy crops, as long as the defined upper scenario limit is not violated. In both scenarios, the actual status quo of national biomass use in 2015 serves as a starting point.  Biomass imports are not allowed in order to avoid a shift of negative environmental effects abroad. For all scenarios, it is assumed that Europe and especially the neighbouring countries of Germany follow similar, ambitious climate targets and that no relocation of industries or imports arise. Carbon capture and storage (CCS) is not considered in this study.

Within the model a discount rate is considered for the investment costs. According to the recommendations of \citet{Steinbach.2015b}, considering the methodology to derive social discount rates as well as discount rates used in analysed energy scenarios, the applied value in this model is set to 4\%.

\section{Results}

\subsection{Scenario results}

\begin{figure*}[ht]
	\centering
		\includegraphics[trim = 40px 0px 40px 5px, width=0.95\textwidth]{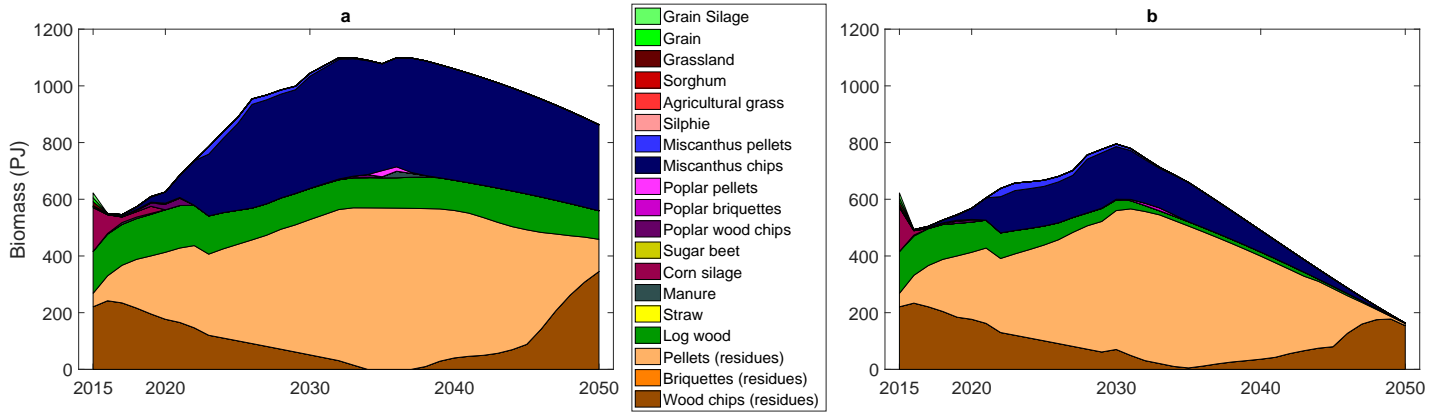}
		\caption{Model resulting consumption of biomass products in case (a) and (b) in a yearly resolution.}
	\label{fig:Result_BiomassProducts}
\end{figure*}

\begin{figure*}[!htp]
	\centering
		\includegraphics[trim = 50px 0px 45px 50px, width=0.81\textwidth]{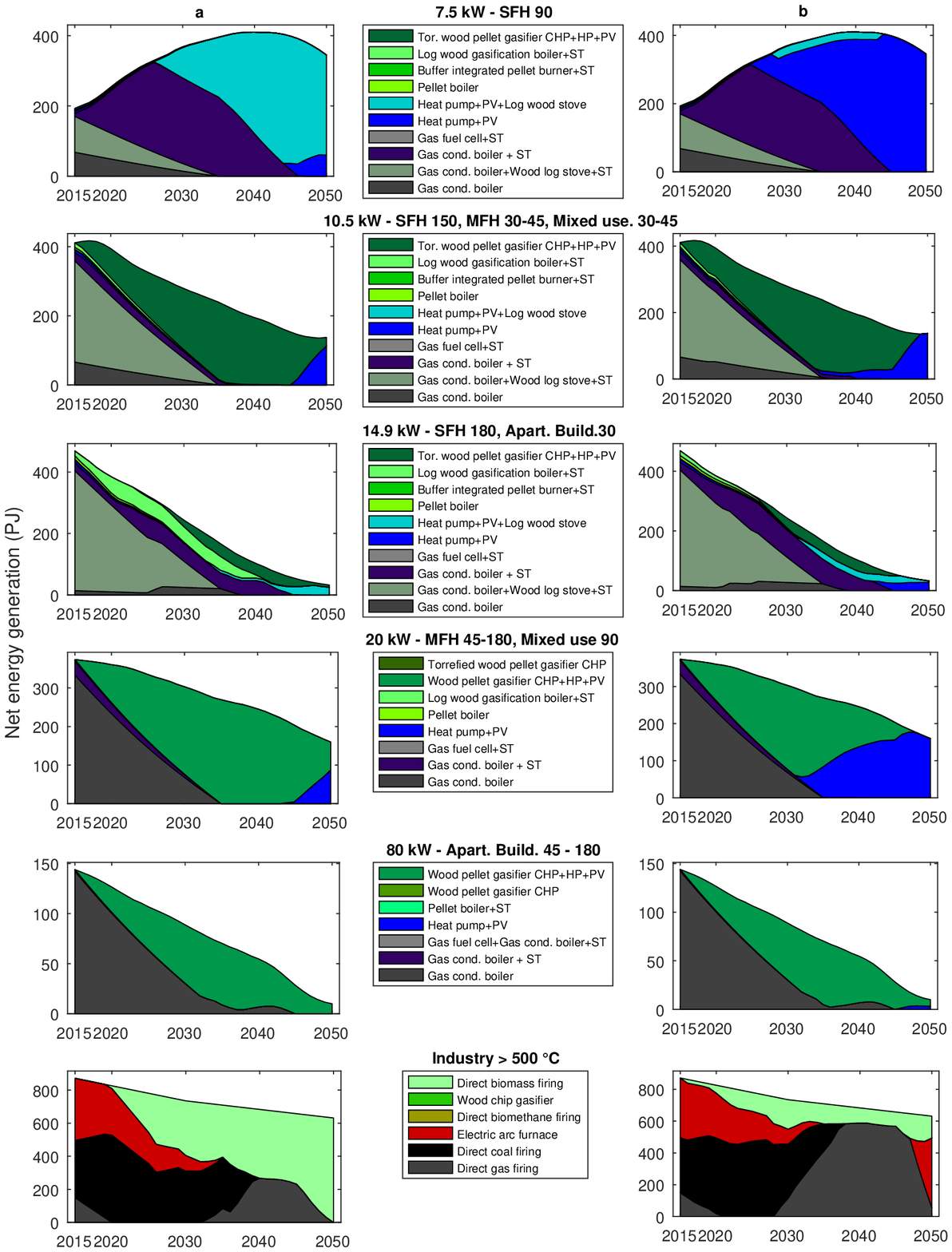}
		\caption{Model resulting development of the technology shares in selected heat sub-sectors in case (a) and (b). The sub-sectors in which biomass technologies are most competitive are illustrated (6 out of 19). SFH = Single Family Houses; MFH = Multi Family Houses; ST = Solar thermal; PV = Photovoltaic; HP = Heat Pump; CHP = Combined Heat and Power}
	\label{fig:Result_SubMarket}
\end{figure*}

In the following paragraph, a transformation path towards a 95\% emission reduction in 2050 in the heat sector is shown. Modelling results are shown for cases (a) and (b) from 2015 to 2050. The market share of all technology types is shown in Fig. \ref{fig:Result_Market}.  As expected, the major market share shifts from natural gas technologies in 2015 to power based heat pumps in 2050. The share of bioenergy in the year 2050 is at 29.0 \% in scenario (a) and 5.7 \% in scenario (b). In both cases, the complete pre-allocated biomass potential is used up from the year 2035 onwards. The largest biomass shares are holding wood chip and pellet technologies. Additionally, in case (a), log wood technologies hold a constant market share of $\sim3 \%$.

A more detailed illustration shows which biomass products are used for heating or CHP technologies, see Fig. \ref{fig:Result_BiomassProducts}. In 2015, one third of the utilised biomass was in the form of biogas, mostly based on corn silage. Without federal subsidies, as it is the case in this model, biogas production is not competitive and market shares decrease rapidly in both scenarios. A constant use of log wood over time is found in case (a), however, log wood technologies are the least cost competitive wood based bioenergy technologies, as their market share decreases rapidly with decreasing biomass potential in case (b) from 2030 onwards. In 2015 residual wood was mainly used for wood chip technologies. The model results show, that in a 95 \% emission reduction scenario the use of residual wood is most competitive over the next three decades in the form of pellets. However, in the last years until 2050, the use of residual wood in the form of wood chips is the favourable option to fulfil climate targets in a cost optimal way.

The available land area for energy crops is cultivated with Miscanthus and processed to chips beginning after the decreasing cultivation of biogas feedstocks, see Fig. \ref{fig:Result_BiomassProducts}. Due to low feedstock costs and high yields, Miscanthus is a competitive option in such a scenario. Notable is the use of Miscanthus in form of chips in contrast to the use of residual wood in form of pellets.

Fig. \ref{fig:Result_SubMarket} shows in which specific sub-sectors and technology concepts the biomass potential is distributed. In six sub-sectors, biomass technologies are competitive options in both scenarios. Five of these sub-sectors belong to the private household sector, in which pellet CHP and torrefied pellet CHP technologies in combination with a heat pump and a photovoltaic system are most competitive over the next three decades. However, between 2040 and 2050, with emission targets to be fulfilled and increasing power prices, a shift of biomass use towards high temperature industry applications is carried out. Consequently, pellet technologies are replaced by heat pumps or log wood technologies after their lifetime expansion.

The market share of log wood technologies is strongly dependent on the available biomass potential, as it is the least competitive wood based option. In case (a), with a high available potential, market shares are constant. Log wood achieves a share of $\sim80\%$ in the 7,5 kW single family houses sector, where the log wood stove is combined with a heat pump and photovoltaic system, while in case (b) this technology holds only a minor market share.  

To sum it up: in the trade and commerce sub-sectors none of the defined bioenergy technologies are a competitive option. Pellet-CHP and  log wood technologies are favourable options in the private household sector, but only in combination with a heat pump and PV-system. Towards 2050, the use of residual wood is more cost efficient in high temperature heat applications.

\section{Discussion}

\begin{figure*}[ht]
	\centering
		\includegraphics[trim = 40px 0px 40px 5px, width=0.95\textwidth]{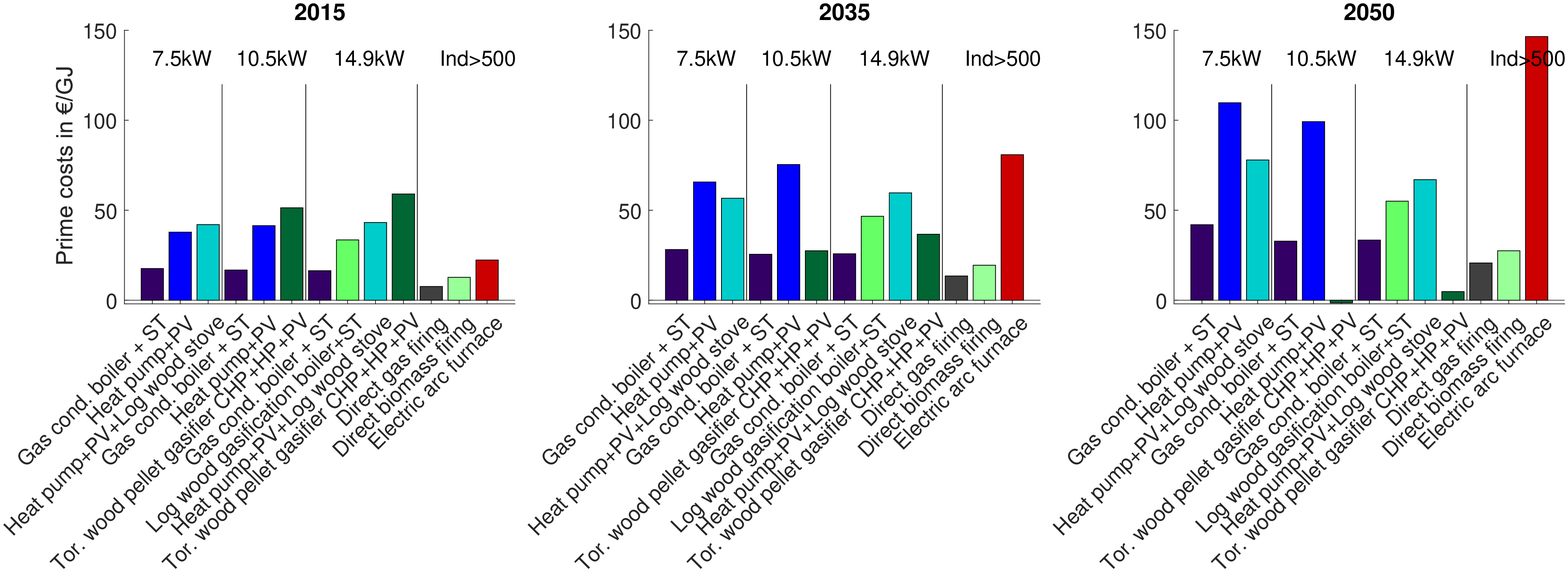}
		\caption{Merit order of the most competitive biomass technologies and their corresponding competitors in selected sub-sectors for the years 2015, 2035 and 2050. Selected sub-sectors are from the private household sector 7.5 kW, 10.5 kW, 14.9 kW and Industry > 500 °C. ST = Solar thermal; PV = Photovoltaic; HP = Heat Pump; CHP = Combined Heat and Power}
	\label{fig:Result_MeritOrder}
\end{figure*}

In this paper, the future role of biomass in a sustainable heat sector is investigated. First of all, the results show that a substantial emission reduction of 95\% compared to 1990 is possible in the German heat sector. A reduction of 98\%, as it is the case in other studies using 'backup capacities' \cite{Repenning.2015,Koch.2018}, was not possible. Second, bioenergy is a competitive option within the defined scenario framework, which confirms the hypothesis from \citet{Thran.2015b,Thran.2016,Thran.2017} expecting to have more competitive applications for wood based biomass in the heat sector compared to the transport and electricity sector. Third, it is identified which biomass products are most competitive in which technology systems and on which sub-sectors of the heat sector.

According to the model results, in the next three decades until 2040-2045 biomass is identified to be most competitive in the private household sector, which is in line with \citet{Koch.2018} and \citet{Repenning.2015}. The most favourable options are decentralised hybrid CHP combustion applications using residual wood as feedstock. Especially the combination of a (torrefied-) wood pellet gasifier CHP with a heat pump and a PV-system is a favourable option. This is a unique finding in energy systems modelling. One reason for this finding is that in available studies on the German energy transition, bioenergy is only considered as single technology option and not analysed in hybrid heat systems \cite{Nitsch.2012,Repenning.2015,Schlesinger.2014,Pfluger.2017,Szarka.2017}. Additionally, this finding shows that the future power price development has a strong impact on the competitiveness of heating systems. Fig. \ref{fig:Result_MeritOrder} shows the merit order of the prime costs for the most competitive biomass options and their corresponding competitors in selected sub-sectors for 2015, 2035 and 2050. With increasing power prices in 2035 and 2050 (see Fig. \ref{fig:PowerGasPrice}), hybrid heat technology systems develop to be the cheapest options of all. Despite these findings, hybrid systems seem to offer the highest degree of self-sufficiency and therefore being more resilient to any kind of feedstock price developments than the competing heating systems. Hence, we conclude that the synergies from hybrid heat technology systems and their GHG mitigation potential are highly underestimated and that such systems can substantially contribute to the success of the energy transition in Germany.

In the long term, in a 95\% reduction scenario, bioenergy is most competitive in high temperature industrial applications in the form of wood chips. From 2040-2045 onwards, biomass use shifts almost entirely from the household sector to high temperature industry applications. This shift away from decentralised private households is in line with \citet{Koch.2018}. The use of wood based biomass for industry applications towards 2050 confirms the projections of several studies (\cite{Gerbert.2018,Szarka.2017,Repenning.2015,Brundinger.2018,acatechDeutscheAkademiederTechnikwissenschaftene.V..2017}). Derived from the results, see Fig. \ref{fig:Result_SubMarket}, we conclude that with emission targets to be fulfilled in 2050 the sub-sector "`Industry > 500 \textdegree C"' requires a major share of renewable technologies. Possible renewable options are heating from biomass or the use of electric arc furnaces. Prime costs of the electric arc are increasing strongly in 2050 compared to biomass heating or heat pumps, see Fig. \ref{fig:Result_MeritOrder}. In the private household sector, the heat pump is an additional option, being more efficient and more cost effective than the electric arcs. Consequently, biomass use shifts to high temperature industry applications, avoiding the use of electric arcs. However, the benefits granted to industry, apart from the generally lower power prices (see Fig \ref{fig:PowerGasPrice}), are not depicted in this model, making the electric arc a possibly cheaper option. On the other hand, the use of electric arcs requires significantly more renewable electricity capacity than the use of heat pumps, which, in contrast, also make use of ambient heat.

In the trade and commerce sector, as well as in district heating, biomass is not a favourable option. For district heating, biogas plants exist today as a result of federal subsidies in the last decades. Without this support, biogas shares are dropping rapidly in case (a) and (b), which is in line with findings from other studies in literature projecting the use of fermentable residues in the transport sector instead of the heat sector, \cite{Koch.2018,Repenning.2015,Schlesinger.2014,Thran.2015b}.  

From the results it is also found that available land for energy crops is cultivated with Miscanthus. Again, this is a unique finding in the modelling of the heat sector. While the cultivation of Miscanthus is an endogenous model result in this study, the above mentioned scenario analysis from literature set the type of energy crops as an input parameter. In addition, it is notable from our results, that Miscanthus is almost exclusively used as chips in industry applications. One explanation is that in private households additional costs for a separator are required if Miscanthus is used in pellet technologies. However, high yields and low production costs lead to a monopoly position among energy crops. So why does Miscanthus play only a minor role in agriculture today? \citet{Witzel.2016} identify several major barriers, e.g. a lack of established markets, high establishment costs as well as uncertainties, arising to a large extent from the necessary long term commitment. These factors are not represented in our optimization model and must be considered separately. Nevertheless, to generate an indicator, a model run excluding perennial crops was performed, resulting in the use of biomethane from maize silage in high temperature industry applications in the long term.

\subparagraph{Limitations:}
\label{sec:Limitations}

Modelling of the heat sector, as it is performed here, depends on several research studies serving as input data. Research insights may change, e.g. the potential of wood based residues was recently corrected downwards \cite{DeutschesBiomasseForschungszentrum.2019}. Do the results and conclusions change, when the pre-allocated biomass potential is changed? How would the results change if the share of the projected district heating network would be higher or if biomass allocation is optimized across all energy sectors? The scenario design with a higher and lower amount of biomass pre-allocated to the heat sector is supposed to represent such shifts of biomass use, but such an approach is limited. However, the outlined results in this study show the same tendency in both scenarios, indicating that these factors might have only a minor impact.

Of course, modelling has its limits, so does this model. The private household sector is depicted in a high level of detail, which was not possible for the industry and district heating sector, due to the limited available data basis. Further research in this direction is highly recommended from the authors' view.

As mentioned before, the power market is not modelled within this study. Therefore a new approach was established for linking the power and heat sector, see section \ref{sec:SectorCoupling}. By setting a scenario framework it is not necessary to have a high temporal resolution, having the advantage of a short model run time leading to the possibility to represent the heat sector and their technology concepts in more detail. To increase the annual resolution to a monthly one seems worthwhile to investigate, since the heat demand, PV yield etc. varies seasonally. However, our model results fit well into the results of the long-term energy scenarios in literature studies \cite{Nitsch.2012,Repenning.2015,Schlesinger.2014,Pfluger.2017,Szarka.2017,Koch.2018}.

When future long-term modelling is done, uncertainties in the input parameters apply and have an effect on the model outcome. Using the applied model, with its short model run time compared to established energy scenario models, opens up the opportunity to apply a comprehensive sensitivity analysis. In future research we will implement all input parameters, having an uncertainty, into a sensitivity analysis and determine the effect of each parameter and all its interactions with all other parameters on the model outcome. A detailed description of the method and results goes beyond the scope of this article.

\section{Conclusions}

In this paper, a 95\% reduction scenario is investigated with two extreme cases of available biomass potential. In both scenarios, the same trends develop, once in an attenuated and once in a stronger manner. It is found that emission targets in the heat sector can be fulfilled in both cases and bioenergy is found to be a future competitive option for heat applications. Especially hybrid heat technology systems were found to be extremely favourable. More specifically, the most cost efficient options for the next decades until 2040 were found to be in the private household sector in form of a hybrid CHP (torrefied-) pellet combustion plant in combination with a heat pump and a PV-system. A key driver for the competitiveness of these systems is the future development of power prices. In times of sector coupling, the advantages of such systems and their potential for emission reduction should not be underestimated and should be taken into account when designing policies. However, in the long term, wood based biomass use is found to shift almost entirely from the private household sector to high temperature applications in the industry. With increasing power prices, the use of wood chips from residues and energy crops in high temperature industry applications is found to be the most cost efficient way to reduce the heat based emissions by 95\% in 2050.

Another finding from this study is, that available land for energy crops is almost entirely cultivated with Miscanthus. Despite several major barriers, arising to a large extent from the long term commitment, this finding should be discussed when designing policies.

\section{Acknowledgements}
\label{sec:Acknowledgements}

Thank you to Öko-Institut e.V. for sharing the heat demand data calculated with B-STar (Building Stock Transformation Model), which have been used in this study for the defined household, trade and commerce and district heating markets \cite{Koch.2018}.

This work was funded by the Helmholtz Association of German Research Centers and supported by Helmholtz Impulse and Networking Fund through Helmholtz Interdisciplinary Graduate School for Environmental Research (HIGRADE). Declarations of interest: none.

\section{Appendix A. Supplementary data}
\label{sec:AppendixA}

Supplementary data related to this article can be found at http://dx.doi.org/10.17632/v2c93n28rj.1

\section{Appendix B.}
\label{sec:AppendixB}

\begin{table*}[htbp]
  \centering
  \caption{Applied heating concepts per sub-sector for private households, trade and commerce. Each row represents a technology concept, each column represents a sub-sector. Per sub-sector the required technology capacity and the specific heat demand of the buildings in kWh/m²a are described. SFH = Single Family House; MFH = Multi Family House; FT = Full Time; PT = Part Time; ST = Solar Thermal; HP = Heat Pump; CHP = Combined Heat and Power; 1: additional peak load heat supply of 25\% of total heat demand from gas condensing boiler; 2: additional peak load heat supply of 20\% of total heat demand from gas condensing boiler}

    \begin{tabular}{r|c|c|c|c|c|c|c|c|c|c|c|c|c|c}
    \toprule
    \multicolumn{1}{r|}{} & \multicolumn{1}{l|}{\begin{sideways}2.5 kW - SFH 30 kWh/m²a\end{sideways}} & \multicolumn{1}{l|}{\begin{sideways}5 kW - SFH 45, MFH 30, Mixed use. 30\end{sideways}} & \multicolumn{1}{l|}{\begin{sideways}7.5 kW - SFH 90\end{sideways}} & \multicolumn{1}{l|}{\begin{sideways}10.5 kW - SFH 150, MFH 30-45, Mixed use. 30-45\end{sideways}} & \multicolumn{1}{l|}{\begin{sideways}14.9 kW - SFH 180, Apart. Build.30\end{sideways}} & \multicolumn{1}{l|}{\begin{sideways}20 kW - MFH 45-180, Mixed use 90\end{sideways}} & \multicolumn{1}{l|}{\begin{sideways}80 kW - Apart. Build. 45 - 180\end{sideways}} & \multicolumn{1}{l|}{\begin{sideways}45 kW - Apart. Build. 45\end{sideways}} & \multicolumn{1}{l|}{\begin{sideways}27 kW - Mixed use \& trade 30-180\end{sideways}} & \multicolumn{1}{l|}{\begin{sideways}31 kW - FT Accommodation since 1984\end{sideways}} & \multicolumn{1}{l|}{\begin{sideways}45 kW - FT Accommodation until 1983\end{sideways}} & \multicolumn{1}{l|}{\begin{sideways}45 kW - PT Accommodation/sport/culture\end{sideways}} & \multicolumn{1}{l|}{\begin{sideways}35 kW - PT Accommodation/sport/culture/trade\end{sideways}} & \multicolumn{1}{l}{\begin{sideways}60 kW - Sport/culture 180\end{sideways}} \\
    \midrule
    Electric direct heating + ST & \multicolumn{1}{l|}{\ding{53}} &       &       &       &       &       &       & \multicolumn{1}{c|}{} &       &       & \multicolumn{1}{c|}{} &       & \multicolumn{1}{c|}{} &  \\
    \midrule
    Gas condensing boiler  &       & \multicolumn{1}{l|}{\ding{53}} & \multicolumn{1}{l|}{\ding{53}} & \multicolumn{1}{l|}{X} & \multicolumn{1}{l|}{\ding{53}} & \multicolumn{1}{l|}{\ding{53}} & \multicolumn{1}{l|}{\ding{53}} & \ding{53}     & \multicolumn{1}{l|}{\ding{53}} & \multicolumn{1}{l|}{\ding{53}} & \ding{53}     & \multicolumn{1}{l|}{\ding{53}} & \ding{53}     & \multicolumn{1}{l}{\ding{53}} \\
    \midrule
    Gas condensing boiler + ST &       & \multicolumn{1}{l|}{\ding{53}} & \multicolumn{1}{l|}{\ding{53}} & \multicolumn{1}{l|}{\ding{53}} & \multicolumn{1}{l|}{\ding{53}} & \multicolumn{1}{l|}{\ding{53}} & \multicolumn{1}{l|}{\ding{53}} & \ding{53}     & \multicolumn{1}{l|}{\ding{53}} & \multicolumn{1}{l|}{\ding{53}} & \ding{53}     & \multicolumn{1}{l|}{\ding{53}} & \ding{53}     & \multicolumn{1}{l}{\ding{53}} \\
    \midrule
    Gas boiler + Log wood stove &       &       & \multicolumn{1}{l|}{\ding{53}} & \multicolumn{1}{l|}{\ding{53}} & \multicolumn{1}{l|}{\ding{53}} &       &       & \multicolumn{1}{c|}{} &       &       & \multicolumn{1}{c|}{} &       & \multicolumn{1}{c|}{} &  \\
    \midrule
    Gas fuel cell + ST &       & \multicolumn{1}{l|}{\ding{53}} & \multicolumn{1}{l|}{\ding{53}} & \multicolumn{1}{l|}{\ding{53}} & \multicolumn{1}{l|}{\ding{53}} & \multicolumn{1}{l|}{\ding{53}} & \multicolumn{1}{l|}{\ding{53}$^{1}$} & \ding{53}$^{2}$    & \multicolumn{1}{l|}{\ding{53}} & \multicolumn{1}{l|}{\ding{53}} & \ding{53}$^{2}$    & \multicolumn{1}{l|}{\ding{53}$^{1}$} & \ding{53}$^{2}$    & \multicolumn{1}{l}{\ding{53}$^{1}$} \\
    \midrule
    Heat pump +  PV & \multicolumn{1}{l|}{\ding{53}} & \multicolumn{1}{l|}{\ding{53}} & \multicolumn{1}{l|}{\ding{53}} & \multicolumn{1}{l|}{\ding{53}} & \multicolumn{1}{l|}{\ding{53}} & \multicolumn{1}{l|}{\ding{53}} & \multicolumn{1}{l|}{\ding{53}} & \multicolumn{1}{c|}{} & \multicolumn{1}{l|}{\ding{53}} & \multicolumn{1}{l|}{\ding{53}} & \multicolumn{1}{c|}{} & \multicolumn{1}{l|}{\ding{53}} & \multicolumn{1}{c|}{} & \multicolumn{1}{l}{\ding{53}} \\
    \midrule
    Heat pump + PV + ST & \multicolumn{1}{l|}{\ding{53}} &       &       &       &       &       &       & \ding{53}     &       &       & \ding{53}     &       & \ding{53}     &  \\
    \midrule
    Heat pump + PV + log wood stove &       &       & \multicolumn{1}{l|}{\ding{53}} & \multicolumn{1}{l|}{\ding{53}} & \multicolumn{1}{l|}{\ding{53}} &       &       & \multicolumn{1}{c|}{} &       &       & \multicolumn{1}{c|}{} &       & \multicolumn{1}{c|}{} &  \\
    \midrule
    Heat pump + PV + Pellet boiler &       &       &       &       &       &       &       & \ding{53}     &       &       & \ding{53}     &       & \ding{53}     & \multicolumn{1}{l}{\ding{53}} \\
    \midrule
    Buffer int. pellet burner + ST & \multicolumn{1}{l|}{\ding{53}} & \multicolumn{1}{l|}{\ding{53}} & \multicolumn{1}{l|}{\ding{53}} & \multicolumn{1}{l|}{\ding{53}} & \multicolumn{1}{l|}{\ding{53}} &       &       & \multicolumn{1}{c|}{} &       &       & \multicolumn{1}{c|}{} &       & \multicolumn{1}{c|}{} &  \\
    \midrule
    Pellet boiler &       & \multicolumn{1}{l|}{\ding{53}} & \multicolumn{1}{l|}{\ding{53}} & \multicolumn{1}{l|}{\ding{53}} & \multicolumn{1}{l|}{\ding{53}} & \multicolumn{1}{l|}{\ding{53}} &       & \multicolumn{1}{c|}{} & \multicolumn{1}{l|}{\ding{53}} & \multicolumn{1}{l|}{\ding{53}} & \multicolumn{1}{c|}{} & \multicolumn{1}{l|}{\ding{53}} & \multicolumn{1}{c|}{} &  \\
    \midrule
    Pellet boiler + ST &       &       &       &       &       &       & \multicolumn{1}{l|}{\ding{53}} & \ding{53}     &       & \multicolumn{1}{l|}{\ding{53}} & \ding{53}     & \multicolumn{1}{l|}{\ding{53}} & \ding{53}     & \multicolumn{1}{l}{\ding{53}} \\
    \midrule
    Log wood stove + ST & \multicolumn{1}{l|}{\ding{53}} &       &       &       &       &       &       & \multicolumn{1}{c|}{} &       &       & \multicolumn{1}{c|}{} &       & \multicolumn{1}{c|}{} &  \\
    \midrule
    Log wood gasification boiler + ST &       & \multicolumn{1}{l|}{\ding{53}} & \multicolumn{1}{l|}{\ding{53}} & \multicolumn{1}{l|}{\ding{53}} & \multicolumn{1}{l|}{\ding{53}} & \multicolumn{1}{l|}{\ding{53}} &       & \multicolumn{1}{c|}{} & \multicolumn{1}{l|}{\ding{53}} &       & \multicolumn{1}{c|}{} &       & \multicolumn{1}{c|}{} &  \\
    \midrule
    Wood chip boiler + ST &       &       &       &       &       &       &       & \multicolumn{1}{c|}{} &       &       & \ding{53}     &       & \multicolumn{1}{c|}{} &  \\
    \midrule
    Torrefied wood pellet gasifier CHP & \multicolumn{1}{l|}{\ding{53}} & \multicolumn{1}{l|}{\ding{53}} &       &       &       & \multicolumn{1}{l|}{\ding{53}} &       & \multicolumn{1}{c|}{} &       &       & \multicolumn{1}{c|}{} &       & \multicolumn{1}{c|}{} &  \\
    \midrule
    Torr. wood pellet g. CHP + HP + PV &       & \multicolumn{1}{l|}{\ding{53}} & \multicolumn{1}{l|}{\ding{53}} & \multicolumn{1}{l|}{\ding{53}} & \multicolumn{1}{l|}{\ding{53}} &       &       & \multicolumn{1}{c|}{} &       &       & \multicolumn{1}{c|}{} &       & \multicolumn{1}{c|}{} &  \\
    \midrule
    Wood pellet gasifier CHP &       &       &       &       &       &       & \multicolumn{1}{l|}{\ding{53}} & \multicolumn{1}{c|}{} &       & \multicolumn{1}{l|}{\ding{53}} & \multicolumn{1}{c|}{} &       & \multicolumn{1}{c|}{} &  \\
    \midrule
    Wood pellet gasifier CHP + HP + PV &       &       &       &       &       & \multicolumn{1}{l|}{\ding{53}} & \multicolumn{1}{l|}{\ding{53}} & \multicolumn{1}{c|}{} & \multicolumn{1}{l|}{\ding{53}} & \multicolumn{1}{l|}{\ding{53}} & \multicolumn{1}{c|}{} & \multicolumn{1}{l|}{\ding{53}} & \multicolumn{1}{c|}{} & \multicolumn{1}{l}{\ding{53}} \\
    \midrule
    Wood pellet gasifier CHP + ST + PV &       &       &       &       &       &       &       & \ding{53}     &       &       & \ding{53}     &       & \ding{53}     &  \\
    \bottomrule
    \end{tabular}%

  \label{tab:Tech2Market}%
\end{table*}%

\clearpage

Details on the defined sub-sectors:

\begin{itemize}
	\item 2.5 kW - SFH 30 kWh/m²a: single or two family house, very well insulated, low temperature heating system, 7 MWh/a heating demand; 2.5 kW thermal peak load, hot water demand 30-40\% of total heat demand
	\item 5 kW - SFH 45, MFH 30, Mixed use. 30: single or two family house well insulated and very well insulated multi-family houses, mixture of low temperature heating system and 70 °C heating, 10.4 MWh/a heating demand; 5 kW thermal peak load, hot water demand 20-24\% of total heat demand
	\item 7.5 kW - SFH 90: single or two family house with 60 to 120 kWh/m²a heat demand, mixture of low temperature heating system with at least 45 °C and 70 °C heating, 14 MWh/a heating demand; 7.5 kW thermal peak load, hot water demand 14-26\% of total heat demand
	\item 10.5 kW - SFH 150, MFH 30-45, Mixed use. 30-45: single or two family house with 120 to 180 kWh/m²a heat demand and well insulated multi-family houses and very well insulated mixed use houses, temperature heating system at least 60 °C, 21 MWh/a heating demand; 10.5 kW thermal peak load, hot water demand 10-40\% of total heat demand
	\item 14.9 kW - SFH 180, Apart. Build.30: single or two family house with more than 180 kWh/m²a heat demand and well insulated multi-family houses and very well insulated big multi family houses, temperature heating system at least 75 °C, 24,5 MWh/a heating demand; 14.9 kW thermal peak load, hot water demand 8-18\% of total heat demand
	\item 20 kW - MFH 45-180, Mixed use 90: mixture of multi family houses and houses with mixed use, temperature of heating system in most cases at least 75 °C, 38 MWh/a heating demand; 20 kW thermal peak load, hot water demand 10-25\% of total heat demand
	\item 80 kW - Apart. Build. 45 - 180: mixture of big multi family houses, temperature of heating system in most cases at least 75 °C or even 85 °C, 165 MWh/a heating demand; 80 kW thermal peak load, hot water demand 18\% of total heat demand
	\item 45 kW - Apart. Build. 45: well insulated multi family houses, temperature of heating system in most cases below 70 °C, 92 MWh/a heating demand; 43 kW thermal peak load, hot water demand 18-29\% of total heat demand
	\item 27 kW - Mixed use \& trade 30-180: mixture of mixed used houses and non-private living buildings, temperature of heating system in most cases at least 75 °C or even 85 °C, 47 MWh/a heating demand; 25 kW thermal peak load, hot water demand 16-19\% of total heat demand
	\item 31 kW - FT Accommodation since 1984: newer non-private living buildings with full day use, temperature of heating system in most cases around 70 °C, 100 MWh/a heating demand; 31 kW thermal peak load, hot water demand 45-50\% of total heat demand
	\item 45 kW - FT Accommodation until 1983: older non-private living buildings with full day use, temperature of heating system in most cases above 75-85 °C, 145 MWh/a heating demand; 45 kW thermal peak load, hot water demand 43-50\% of total heat demand
	\item 45 kW - PT Accommodation/sport/culture: older non-private living buildings with half day use and newer special buildings, temperature of heating system at least 50-60 °C sometimes significantly above that values, 74 MWh/a heating demand; 45 kW thermal peak load, hot water demand 13-16\% of total heat demand
	\item 35 kW - PT Accommodation/sport/culture/trade: mainly older non-private living buildings with half day use and old special buildings both with high specific heating demand, temperature of heating system at least 75-85 °C, 56 MWh/a heating demand; 34 kW thermal peak load, hot water demand 15-19\% of total heat demand
	\item 60 kW - Sport/culture 180: old special buildings with more than 180 kWh/m²a heating demand, temperature of heating system at least 75-85 °C, 100 MWh/a heating demand; 60 kW thermal peak load, hot water demand 13\% of total heat demand
\end{itemize}

\begin{table*}[htbp]
  \centering
  \caption{Applied heating concepts per sub-sector in industry and district heating. ST = Solar Thermal; CHP = Combined Heat and Power; HT = High Temperature}

\begin{tabular}{r|c|c|c|c|c}
\toprule
      & \multicolumn{1}{r|}{\begin{sideways}Industry < 200 °C\end{sideways}} & \multicolumn{1}{r|}{\begin{sideways}Industry 200 - 500 °C\end{sideways}} & \multicolumn{1}{r|}{\begin{sideways}Industry 500 - 1.500 °C\end{sideways}} & \multicolumn{1}{r|}{\begin{sideways}Special coal demand\end{sideways}} & \multicolumn{1}{r}{\begin{sideways}District heating\end{sideways}} \\
\midrule
Gas condensing boiler & \ding{53}     & \ding{53}     &       &       &  \\
\midrule
Gas fuel cell & \ding{53}     &       &       &       &  \\
\midrule
HT heat pump + ST (5\%) & \ding{53}     &       &       &       &  \\
\midrule
Wood chip boiler & X     & \ding{53}     &       &       &  \\
\midrule
Wood chip gasifier CHP & \ding{53}     &       &       &       & \ding{53} \\
\midrule
Heat pump + ST (5\%) + Wood chip boiler (40\%) & \ding{53}     &       &       &       &  \\
\midrule
Gas turbine CHP &       & \ding{53}     &       &       &  \\
\midrule
Biomethane gas turbine CHP &       & \ding{53}     &       &       &  \\
\midrule
Wood chip gasifier with gas turbine CHP &       & \ding{53}     &       &       &  \\
\midrule
Direct gas firing &       &       & \ding{53}     &       &  \\
\midrule
Direct coal firing &       &       & \ding{53}     &       &  \\
\midrule
Electric arc furnace &       &       & \ding{53}     &       &  \\
\midrule
Direct biomethane firing &       &       & \ding{53}     &       &  \\
\midrule
Wood chip gasifier with direct gas firing &       &       & \ding{53}     &       &  \\
\midrule
Direct biomass firing &       &       & \ding{53}     &       &  \\
\midrule
Coke &       &       &       & \ding{53}     &  \\
\midrule
Bio-coke &       &       &       & \ding{53}     &  \\
\midrule
Coal CHP plant &       &       &       &       & \ding{53} \\
\midrule
Gas and steam turbine CHP &       &       &       &       & \ding{53} \\
\midrule
Coal CHP plant with 5\% wood chips &       &       &       &       & \ding{53} \\
\midrule
HT heat pump + ST + Methane CHP boiler &       &       &       &       & \ding{53} \\
\midrule
Waste CHP plant + Wood chip boiler &       &       &       &       & \ding{53} \\
\bottomrule
\end{tabular}%

  \label{tab:Tech2MarketIndustry}%
\end{table*}%

\begin{table*}[tbp]
  \centering
  \caption{Defined application possibilities of the feedstocks in the technologies. CHP = Combined Heat and Power}
    \begin{tabular}{r|c|c|c|c|c|c|c|c|c|c|c|c|c|c|c|c|c}
          & \multicolumn{1}{l|}{\begin{sideways}Gas condensing boiler / fuel cell /plant\end{sideways}} & \multicolumn{1}{l|}{\begin{sideways}Log wood stove\end{sideways}} & \multicolumn{1}{l|}{\begin{sideways}Log wood gasification boiler\end{sideways}} & \multicolumn{1}{l|}{\begin{sideways}Wood pellet boiler/gasifier\end{sideways}} & \multicolumn{1}{l|}{\begin{sideways}Wood pellet CHP\end{sideways}} & \multicolumn{1}{l|}{\begin{sideways}Torrefied wood pellet CHP\end{sideways}} & \multicolumn{1}{l|}{\begin{sideways}Wood chip boiler\end{sideways}} & \multicolumn{1}{l|}{\begin{sideways}Hard coal CHP / coal coke\end{sideways}} & \multicolumn{1}{l|}{\begin{sideways}Wood chip - hard coal CHP\end{sideways}} & \multicolumn{1}{l|}{\begin{sideways}Biomethane applications\end{sideways}} & \multicolumn{1}{l|}{\begin{sideways}Waste CHP plant\end{sideways}} & \multicolumn{1}{l|}{\begin{sideways}Wood chip gasifier CHP\end{sideways}} & \multicolumn{1}{l|}{\begin{sideways}Gas turbine / direct heating\end{sideways}} & \multicolumn{1}{l|}{\begin{sideways}Wood gasifier gasturbine\end{sideways}} & \multicolumn{1}{l|}{\begin{sideways}Coal direct heating\end{sideways}} & \multicolumn{1}{l|}{\begin{sideways}Biomass direct heating\end{sideways}} & \multicolumn{1}{l}{\begin{sideways}Bio-Coke\end{sideways}} \\
    \midrule
    Wood chips (residues) &       &       &       &       &       &       & \ding{53}     &       & \ding{53}     &       & \ding{53}     & \ding{53}     &       & \ding{53}     &       & \ding{53}     & \ding{53} \\
    \midrule
    Briquettes (residues) &       & \ding{53}     &       &       &       &       &       &       &       &       &       &       &       &       &       &       &  \\
    \midrule
    Pellets (residues) &       &       &       & \ding{53}     & \ding{53}     & \ding{53}     &       &       &       &       &       &       &       &       &       & \ding{53}     &  \\
    \midrule
    Log wood &       & \ding{53}     & \ding{53}     &       &       &       &       &       &       &       &       &       &       &       &       &       &  \\
    \midrule
    Straw & \ding{53}     &       &       &       &       &       &       &       &       & \ding{53}     &       &       &       &       &       &       &  \\
    \midrule
    Manure & \ding{53}     &       &       &       &       &       &       &       &       & \ding{53}     &       &       &       &       &       &       &  \\
    \midrule
    Corn silage & \ding{53}     &       &       &       &       &       &       &       &       & \ding{53}     &       &       &       &       &       &       &  \\
    \midrule
    Sugar beet & \ding{53}     &       &       &       &       &       &       &       &       & \ding{53}     &       &       &       &       &       &       &  \\
    \midrule
    Poplar wood chips &       &       &       &       &       &       & \ding{53}     &       & \ding{53}     &       & \ding{53}     & \ding{53}     &       & \ding{53}     &       & \ding{53}     &  \\
    \midrule
    Poplar briquettes &       & \ding{53}     &       &       &       &       &       &       &       &       &       &       &       &       &       &       &  \\
    \midrule
    Poplar pellets &       &       &       & \ding{53}     & \ding{53}     &       &       &       &       &       &       &       &       &       &       & \ding{53}     &  \\
    \midrule
    Miscanthus chips &       &       &       &       &       &       & \ding{53}     &       &       &       & \ding{53}     &       &       &       &       & \ding{53}     &  \\
    \midrule
    Miscanthus briquettes &       &       &       &       &       &       &       &       &       &       &       &       &       &       &       &       &  \\
    \midrule
    Miscanthus pellets &       &       &       & \ding{53}     &       &       &       &       &       &       &       &       &       &       &       & \ding{53}     &  \\
    \midrule
    Silphie & \ding{53}     &       &       &       &       &       &       &       &       & \ding{53}     &       &       &       &       &       &       &  \\
    \midrule
    Agricultural grass & \ding{53}     &       &       &       &       &       &       &       &       & \ding{53}     &       &       &       &       &       &       &  \\
    \midrule
    Sorghum & \ding{53}     &       &       &       &       &       &       &       &       & \ding{53}     &       &       &       &       &       &       &  \\
    \midrule
    Grassland & \ding{53}     &       &       &       &       &       &       &       &       & \ding{53}     &       &       &       &       &       &       &  \\
    \midrule
    Grain & \ding{53}     &       &       &       &       &       &       &       &       & \ding{53}     &       &       &       &       &       &       &  \\
    \midrule
    Grain Silage & \ding{53}     &       &       &       &       &       &       &       &       & \ding{53}     &       &       &       &       &       &       &  \\
    \midrule
    Natural gas & \ding{53}     &       &       &       &       &       &       &       &       &       &       &       & \ding{53}     &       &       &       &  \\
    \midrule
    Coal  &       &       &       &       &       &       &       & \ding{53}     & \ding{53}     &       &       &       &       &       & \ding{53}     &       &  \\
    \midrule
    Plastic waste &       &       &       &       &       &       &       &       &       &       & \ding{53}     &       &       &       &       &       &  \\
    \bottomrule
    \end{tabular}%
  \label{tab:BiomassProducts2Tech}%
\end{table*}%

\begin{table}[t]
  \centering
  \caption{Applied emission factors caused by infrastructure expenses in 2015 \cite{Swisscentreforlifecycleinventories.2010,Swisscentreforlifecycleinventories.2016,Umweltbundesamt.2019} and the calculated allocation factor according the finnish method. The allocation factor is also applied to the deployed feedstock. Infrastructure emissions are linearly reduced by 80\% until 2050. CHP = Combined Heat and Power; PH = Private Household}
\begin{tabular}{rcc}
\toprule
			& \rotatebox{90}{\parbox{10.5em}{Infrastructure emissions in $gCO_{2}-eq/MJ_{out}$}} &  \rotatebox{90}{Allocation factor}\\
\midrule
Electric direct heating & 0.75  &  \\
Gas condensing boiler & 0.25  &  \\
Solar thermal & 6.89  &  \\
Gas fuel cell 125kWe & 5.53  & 0.30 \\
Heat pump & 1.87  &  \\
Wood pellet boiler & 1.72  &  \\
Log wood gasification boiler & 0.55  &  \\
Torrefied wood pellet gasifier & 0.55  &  \\
Buffer integrated pellet burner & 0.55  &  \\
Wood pellet gasifier CHP & 1.93  & 0.59 \\
Gas condensing boiler (Industry) & 0.03  &  \\
Wood chip boiler (PH) & 0.22  &  \\
Wood chip boiler (Industry) & 1.60  &  \\
Wood chip gasifier CHP (Ind. low Temp.) & 0.14  & 0.29 \\
Gas Fuel cell (Industry) & 5.53  & 0.46 \\
High temperature heat pump & 1.94  &  \\
Wood chip gasifier CHP (District heating) & 1.27  & 0.45 \\
Gas turbine CHP & 0.11  & 0.13 \\
Biomethane gas turbine CHP & 0.11  & 0.13 \\
Wood chip gasifier CHP (Ind. high Temp.) & 0.30  & 0.13 \\
Direct Gas firing & 0.03  &  \\
Direct Coal firing & 0.03  &  \\
Electric arc furnace & 0.08  &  \\
Direct biomethane firing & 0.03  &  \\
Wood chip gasifier with direct gas firing & 0.03  &  \\
Direct biomass firing & 0.03  &  \\
Coal CHP plant & 0.11  & 0.13 \\
Gas and steam turbine CHP & 0.13  & 0.34 \\
Coal CHP plant with 5\% wood chips & 0.11  & 0.13 \\
Methane CHP boiler & 1.14  & 0.38 \\
Waste CHP plant & 0.11  & 0.64 \\
Photovoltaic system (gCO$_2$-eq/kWel) & 78.99 &  \\
\bottomrule
\end{tabular}%

  \label{tab:EmissionFactor}%
\end{table}%

\begin{table}[t]
  \centering
  \caption{Applied feedstock emission factors \cite{Swisscentreforlifecycleinventories.2010,Swisscentreforlifecycleinventories.2016,Umweltbundesamt.2019}. Emissions based on power consumed from the grid are calculated according the scenario depended, power mix specific emission factor \cite{Repenning.2015}. In relation to biomass emissions: Including the effects on carbon storage in vegetation and soil, biomass can only be considered CO$_{2}$ neutral if it would rot quickly without energy use (residual and waste materials), or if land and vegetation are managed in such a way that they absorb more CO$_{2}$ than they would without bioenergy use (taking into account indirect land use effects). One example is the establishment of short rotation plantations on pasture land \cite{Byfield.2018}.}
    \begin{tabular}{rr}
    \toprule
					& \rotatebox{90}{\parbox{9em}{Feedstock emissions in $gCO_{2}-eq/MJ_{in}$}} \\
    \midrule
    Wood chips (residues) & 1.36 \\
    Briquettes (residues) & 7.94 \\
    Pellets (residues) & 7.94 \\
    Log wood & 4.47 \\
    Straw & 3.93 \\
    Manure & 0.00 \\
    Corn silage & 7.35 \\
    Sugar beet & 7.20 \\
    Poplar wood chips & 3.83 \\
    Poplar briquettes & 8.25 \\
    Poplar pellets & 8.25 \\
    Miscanthus chips & 4.10 \\
    Miscanthus briquettes & 8.53 \\
    Miscanthus pellets & 8.53 \\
    Silphie & 5.27 \\
    Agricultural grass & 14.83 \\
    Sorghum & 16.11 \\
    Grassland & 15.41 \\
    Grain & 4.78 \\
    Grain Silage & 12.07 \\
    Natural gas & 59.60 \\
    Coal  & 108.00 \\
    Plastic waste & 59.75 \\
    Coal coke & 123.00 \\
    Bio-coke & 27.78 \\
    \bottomrule
    \end{tabular}%
  \label{tab:FeedstockEmission}%
\end{table}%

\begin{table*}[htbp]
  \centering
  \caption{Yield of the defined energy crops \cite{KuratoriumfurTechnikundBauweseninderLandwirtschafte.V..2012} and their corresponding land use in 2015 for heat or combined heat and power applications \cite{Becker.2018b}. SRC = Short Rotation Coppice}
\begin{tabular}{rcc}
\toprule
      & \multicolumn{1}{c}{\multirow{2}[2]{*}{\textbf{Yield\newline{}(GJ/ha)}}} & \textbf{Land use (ha)} \\
      &       & \textbf{2015} \\
\midrule
\textbf{Corn silage} & 177   & 872 000 \\
\textbf{Sugar beet} & 150   & 15 600 \\
\textbf{Grain} & 91    & 151 000 \\
\textbf{Grain Silage} & 138   & 123 000 \\
\textbf{Agr. grass} & 137   & 20 150 \\
\textbf{Grassland} & 90    & 157 849 \\
\textbf{Silphie} & 126   & 400 \\
\textbf{Sorghum} & 152   & 0 (est.) \\
\textbf{SRC} & 137   & 6 630 \\
\textbf{Miscanthus} & 273   & 4 500 \\
\bottomrule
\end{tabular}%

  \label{tab:YieldLandUse}%
\end{table*}%

\begin{table*}[htbp]
  \centering
  \caption{Applied surcharges in the model based on own calculations.}
	\begin{tabular}{rc}
	\toprule
				& \textbf{Surcharge (\euro{}/GJ)} \\
	\midrule
	\textbf{Pellets compared to wood chips} & 5 \\
	\textbf{Pellet torrefication} & + 14 \% \\
	\textbf{Briquettes compared to wood chips} & 7 \\
	\textbf{Separator for torrefied poplar pellets in pellet technologies} & 0.3 \\
	\textbf{Separator for miscanthus pellets in pellet technologies} & 0.2 \\
	\textbf{Separator for poplar briquettes in log wood technologies} & 0.05 \\
	\textbf{Separator for straw in wood chip technologies} & 0.4 \\
	\textbf{Separator for poplar wood chips in wood chip gasification technologies} & 0.2 \\
	\textbf{Separator for miscanthus chips in wood chip technologies} & 0.2 \\
	\textbf{Transport fee for wood based feedstocks per delivery} & 50 \euro{} \\
	\bottomrule
	\end{tabular}%

  \label{tab:surcharges}%
\end{table*}%

\clearpage

\section*{References}

\bibliography{mybibfile}

\end{document}